\documentclass[sigconf, screen,nonacm]{acmart}
\makeatletter
\newcommand*\dashline{\rotatebox[origin=c]{90}{$\dabar@\dabar@$}}
\makeatother
\AtBeginDocument{%
  }

\setcopyright{none}
\acmISBN{978-1-4503-XXXX-X/2025/09}

\usepackage{fontawesome5}
\usepackage{subcaption}

\begin{document}

\title{Exploring the Effect of Context-Awareness and Popularity Calibration on Popularity Bias in POI Recommendations}


\author{Andrea Forster}
\orcid{0009-0008-6818-1916}
\affiliation{%
  \institution{Graz University of Technology}
  \city{Graz}
  \country{Austria}}
\email{andrea.forster@student.tugraz.at}

\author{Simone Kopeinik}
\orcid{0000-0002-6440-7286}
\affiliation{%
  \institution{Know Center Research GmbH}
  \city{Graz}
  \country{Austria}}
\email{skopeinik@know-center.at}

\author{Denis Helic}
\orcid{0000-0003-0725-7450}
\affiliation{%
  \institution{Graz University of Technology}
  \city{Graz}
  \country{Austria}}
\email{dhelic@tugraz.at}

\author{Stefan Thalmann}
\orcid{0000-0001-6529-7958}
\affiliation{%
  \institution{BANDAS Center}
  \institution{University of Graz}
  \city{Graz}
  \country{Austria}}
\email{stefan.thalmann@uni-graz.at}

\author{Dominik Kowald}
\orcid{0000-0003-3230-6234}
\affiliation{%
  \institution{Know Center Research GmbH}
  \city{Graz}
  \country{Austria}\\
  \institution{Graz University of Technology}
  \city{Graz}
  \country{Austria}}
\email{dkowald@know-center.at}

%\renewcommand{\shortauthors}{Forster, et al.}

\begin{abstract}
Point-of-interest (POI) recommender systems help users discover relevant locations, but their effectiveness is often compromised by popularity bias, which disadvantages less popular, yet potentially meaningful places. This paper addresses this challenge by evaluating the effectiveness of context-aware models and calibrated popularity techniques as strategies for mitigating popularity bias. Using four real-world POI datasets (Brightkite, Foursquare, Gowalla, and Yelp), we analyze the individual and combined effects of these approaches on recommendation accuracy and popularity bias. Our results reveal that context-aware models cannot be considered a uniform solution, as the models studied exhibit divergent impacts on accuracy and bias. In contrast, calibration techniques can effectively align recommendation popularity with user preferences, provided there is a careful balance between accuracy and bias mitigation. Notably, the combination of calibration and context-awareness yields recommendations that balance accuracy and close alignment with the users' popularity profiles, i.e., popularity calibration.
\end{abstract}




%\received{29 April 2025}

\maketitle

\begingroup
\renewcommand\thefootnote{}\footnotetext{%
  \hspace{-1.5em}\raisebox{5pt}{%
    \begin{minipage}[t]{\columnwidth}
      \footnotesize
      © Andrea Forster, Simone Kopeinik, Denis Helic, Stefan Thalmann, and Dominik Kowald, 2025. This is the author's version of the work entitled ``Exploring the Effect of Context-Awareness and Popularity Calibration on Popularity Bias in POI Recommendations. It is posted here for your personal use. Not for redistribution. The definitive version of record was accepted for publication in the \textit{19th ACM Conference on Recommender Systems (RecSys 2025)}, \url{https://doi.org/10.1145/3705328.3748017}
    \end{minipage}%
  }%
}
\endgroup

\section{Introduction}

Tourism provides a rich ground for recommender systems (RS), supporting tasks such as destination planning, hotel, transport, or points-of-interest (POI) selection~\citep{sarkar_tourism_2023}. POI recommendations are particularly challenging, characterized by datasets with high sparsity and the reliance on observational data rather than explicit rating data to infer user preferences~\citep{becker2015, sanchez_point_2022}. Additional complexity is added by the use of contextual features such as geographic proximity, temporal dynamics, or social network connections to improve personalization~\citep{sarkar_tourism_2023,lacic2014towards,lacic2014recommending,sanchez_point_2022}. More recently, social media data, such as geotagged images, have also been explored to enhance POI recommendations~\citep{sarkar_tourism_2023}. 
A growing body of work highlights fairness and sustainability concerns in tourism and location-based RS. In this regard, popularity bias~\citep{abdollahpouri_multi-sided_2020,lacic2022drives,kowald2022popularity} poses a significant challenge, where popular POI dominate recommendations, limiting exposure to diverse and lesser-known places, attributed to the sensitivity of many well-known RS (especially collaborative filtering) to popularity indications~\citep{deldjoo_explaining_2021,rahmani_unfairness_2022}. Popularity bias can lead to unfair treatment of both item providers, where niche locations receive less exposure~\citep{rahmani_unfairness_2022,abdollahpouri_multi-stakeholder_2019,banerjee_review_2023}, and consumers (i.e., RS users), not catering to those who prefer less mainstream locations~\citep{rahmani_unfairness_2022,kowald_unfairness_2020,chen_bias_2020,abdollahpouri_multi-sided_2020}. In tourism, this can elevate overcrowding, environmental degradation, or reduced user satisfaction~\citep{rahmani_unfairness_2022,abdollahpouri_multi-stakeholder_2019,banerjee_review_2023}.

Although prior research highlights the potential of context-awareness~\citep{banerjee_review_2023, rahmani_unfairness_2022} and calibrated popularity (CP)~\citep{abdollahpouri_user-centered_2021, lesota_exploring_2022,klimashevskaia_evaluating_2023,ungruh_putting_2024} as effective strategies to mitigate popularity bias, these two approaches have so far been studied in isolation. To our knowledge, CP has not yet been applied in the context of POI recommendation, nor combined with context-aware models. This raises questions about whether these strategies complement each other and how they jointly affect the trade-off between accuracy and popularity bias.

In our work, we address this research gap by systematically evaluating CP and two well-known context-aware models, a \textit{LO}cation \textit{RE}commendation approach (LORE)~\citep{zhang_lore_2014} and a collaborative filtering model that harnesses \textit{U}ser preference, \textit{S}ocial influence, and \textit{G}eographical influence (USG)~\citep{ye_exploiting_2011}, independently and in combination, and comparing them to the non-contextualized baseline Bayesian Personalized Ranking (BPR)~\citep{rendle_bpr_2012}, thus contributing empirical insight into how context-aware RS and calibration interact, and how these methods affect different user groups. Using four real-world POI datasets (Brightkite, Foursquare, Gowalla, and Yelp), we group users based on their preference for popular locations ($LowPop$, $MedPop$, and $HighPop$), and analyze how the RS approaches impact accuracy and popularity within these groups. We formulate two research questions to guide this work and to structure our methodology as well as results presentation:

\begin{itemize}
    \item 
    \textbf{RQ1.} To what extent can context-aware recommendations (LORE, USG) and calibration-based debiasing (CP) individually mitigate popularity bias in POI recommendations, and how does this impact accuracy, compared to a non-contextual baseline (BPR)?
    \item
    \textbf{RQ2.} Does the combination of context-aware POI recommendations and calibration-based debiasing (CP) improve the trade-off between recommendation accuracy and popularity bias, compared to their respective purely context-aware versions (LORE, USG)?
\end{itemize}

%\vspace{1mm} \noindent
Our results show that non-contextualized models like BPR disadvantage $LowPop$ users, while the performance of context-aware models depends predominantly on the model and dataset. CP helps align recommendations with user preferences, but may affect accuracy (RQ1). Combining CP with context-aware models yields interesting results: USG combined with CP achieves higher accuracy, but retains more popularity bias, while LORE combined with CP increases the popularity of the LORE base version, yet offers the closest match to user popularity profiles out of all methods and combinations studied in our work (RQ2). 

Our findings provide valuable insights for researchers and practitioners working on mitigating popularity bias in POI recommendation systems and can inform the design of user studies aimed at evaluating user experience.

%%%%%%%%%%%%%%%%%%%%%%%%%%%%%%%%
\section{Background and Method}
\subsection{Overview of Related Work}
Several studies explore how the effects of popularity bias and overcrowding can be mitigated in RS: 
\citet{ghanem_balancing_2022} model trade-offs between consumer utility and provider profit; \citet{merinov_sustainability_2022} optimize travel itineraries to reduce POI congestion;
\citet{banerjee_modeling_2025} propose a fairness score incorporating environmental and seasonal constraints and location popularity into recommendations; 
\citet{massimo_popularity_2021} explore the trade-offs between accuracy and novelty in POI recommendation, concluding that users prioritize precision over novelty and struggle to assess unknown suggestions.
\citet{rahmani_role_2022,rahmani_unfairness_2022, banerjee_review_2023} find context-aware POI recommendation models that incorporate geographical, temporal, social, or categorical information to produce more diverse recommendation lists than traditional RS (e.g., collaborative filtering). \citet{abdollahpouri_user-centered_2021} introduce CP, a user-centered debiasing method that aligns the distribution of head, mid, and tail (H, M, T) items in recommendation lists with users’ historical interaction patterns. \citet{klimashevskaia_evaluating_2023} evaluate CP in a movie streaming setting, finding that it improves fairness without degrading performance; \citet{ungruh_putting_2024} conducted a user study in the music domain showing that a mix of familiar and unfamiliar items leads to more satisfactory recommendations; and \citet{lesota_exploring_2022} introduce group-specific trade-off parameters to improve fairness across user subgroups, calling for more targeted criteria to select mitigation parameters.

\subsection{Datasets}
In our work, we leverage four datasets commonly used in tourism and POI recommendation research~\citep{zhou_improved_2024, tourani_capri_2024,sanchez_point_2022, rahmani_unfairness_2022,banerjee_review_2023}: Foursquare\footnote{\url{https://www.kaggle.com/datasets/chetanism/foursquare-nyc-and-tokyo-checkin-dataset}}, Yelp\footnote{\url{https://www.yelp.com/dataset}}, Brightkite\footnote{\url{https://snap.stanford.edu/data/loc-brightkite.html}}, and Gowalla\footnote{\url{https://snap.stanford.edu/data/loc-gowalla.html}}. Due to limited space, we illustrate results only for Foursquare and Yelp, but provide them for all four datasets on GitHub\footnote{\url{https://github.com/andreafooo/POI_RS_PopBias_Mitigation}}. 
Each dataset includes \textit{user ID}, \textit{timestamp}, \textit{check-in} location, and POI coordinates. To reduce the computational costs of our study, we create data samples. The Foursquare sample includes 1,500 users, 2,804 items, and 69,401 unique check-ins (sparsity = 98.4\%). Our Yelp sample includes 1,500 users, 4,515 items, and 35,288 unique check-ins (sparsity = 99.5\%), as illustrated in Table~\ref{tab:data_overview}. We group users by their average profile popularity, based on normalized location check-in frequencies~\citep{abdollahpouri_unfairness_2019, kowald_unfairness_2020}: the bottom 20\% ($LowPop$), middle 60\% ($MedPop$), top 20\% ($HighPop$). We similarly classify items as tail (T, bottom 20\%), mid (M, middle 60\%), and head (H, top 20\%)~\citep{abdollahpouri_user-centered_2021,ungruh_putting_2024}. Finally, we apply a user-based temporal split \citep{sanchez_point_2022} to create training (65\%), validation (15\%), and test (20\% most recent check-ins) sets.
%, where multiple check-ins are transformed into a check-in count. 

\begin{table}[ht]
\small % or \scriptsize if you need it even tighter
\centering
\caption{Descriptive statistics of the datasets used in this work. For the sake of space, we only report results for Foursquare and Yelp in the remainder of this paper.}
\label{tab:data_overview}
\begin{tabular}{lccccc}
\toprule
Dataset     & Users & Items & Unique check-ins & Sparsity \\
\midrule
Brightkite  & 600 & 794   &  15,341 & 0.967798 \\

Foursquare & 1,500 & 2,804  &  69,401 & 0.983500 \\

Gowalla    & 1,500 & 7,579  &  53,679    & 0.995278 \\

Yelp       & 1,500 & 4,515  &   35,288  & 0.994790 \\
\bottomrule
\end{tabular}
\end{table}

\subsection{Popularity Bias Mitigation}
\paragraph{\textbf{Context-Aware POI Recommendation.}} We use two well-known POI recommendation models, which are implemented in the CAPRI framework\footnote{\url{https://github.com/CapriRecSys/CAPRI}}. The first model, LORE, integrates sequential, social, and geographical influences by combining additive Markov chains with a location-location transition graph to model user movement patterns and POI transitions~\citep{zhang_lore_2014}. The second model, USG, is a hybrid model that combines user- and friend-based collaborative filtering with geographical influence using a naive Bayes approach~\citep{ye_exploiting_2011}. We do not use data on social relations in our models and experiments, since not all of our datasets contain such information.

\paragraph{\textbf{Calibrated Popularity (CP)}} CP is a re-ranking method that adjusts a base recommendation list to better reflect a user’s historical preferences for item popularity levels. Proposed by \citet{abdollahpouri_user-centered_2021} and based on the idea of calibrated recommendations by \citet{steck_calibrated_2018}, CP selects a refined recommendation list $L^*_u$ of size $n$ from a larger base list $L_u$ of size $m$, using a weighted optimization that balances item relevance and distributional similarity.

\begin{table*}[t!]
\centering
\caption{RQ1. Performance of context-aware recommendation models and CP applied on BPR in relation to the BPR baseline. Metrics include nDCG, ARP, and PopLift. Symbols indicate the preferred direction for each metric: $\downarrow$ (lower is better), $\uparrow$ (higher is better), and $\rightarrow 0$ (closer to zero is better), and best values are shown in bold. For BPR, absolute values are shown; $\Delta$\% values for LORE, USG $CP_H$ and $CP_\Im$. Significant relations are indicated by ** via t-test ($p < 0.05$), Bonferroni-corrected for each metric.}
    \label{tab:rq1}
    \resizebox{\textwidth}{!}{%
    \begin{tabular}{l|llllll|llllll|llllll}
\toprule
\multicolumn{1}{c}{\textbf{Group}} & \multicolumn{6}{c}{\textbf{nDCG $\uparrow$ | $\Delta$\% nDCG}} & \multicolumn{6}{c}{\textbf{ARP $\downarrow$ | $\Delta$\% ARP}}  & \multicolumn{6}{c}{\textbf{PopLift $\rightarrow 0$ | $\Delta$\% PopLift}} \\
 & BPR & \dashline & LORE & USG & $CP_H$ & $CP_\Im$ & BPR & \dashline & LORE & USG & $CP_H$ & $CP_\Im$ & BPR & \dashline & LORE & USG & $CP_H$ & $CP_\Im$ \\
 \hline
\midrule
\multicolumn{19}{c}{\textbf{Foursquare}} \\
$LowPop$ & 0.0395 & \dashline & -56.62\%** & -43.28\%** & \textbf{+0.54\%} & -21.28\% & 0.0795 & \dashline & \textbf{-91.30\%**} & +94.71\%** & -0.08\% & -30.97\%** & 4.3299 & \dashline & \textbf{-110.89\%**} & +149.05\%** & -0.11\% & -39.98\%** \\
$MedPop$ & \textbf{0.1084} & \dashline & -88.77\%** & -11.81\%** & -0.04\% & -14.92\%** & 0.1009 & \dashline & \textbf{-93.69\%**} & +32.79\%** & -0.41\%** & -17.69\%** & 2.2977 & \dashline & \textbf{-134.47\%**} & +48.68\%** & -0.72\%** & -26.28\%** \\
$HighPop$ & \textbf{0.1655} & \dashline & -96.63\%** & -2.11\% & -0.14\% & -6.43\%** & 0.1089 & \dashline & \textbf{-94.31\%**} & +18.09\%** & -0.37\%** & -6.85\%** & 1.2302 & \dashline & \textbf{-171.04\%**} & +33.50\%** & -0.82\%** & -13.57\%** \\
$All$ & \textbf{0.1060} & \dashline & -88.83\%** & -11.13\%** & -0.03\% & -12.74\%** & 0.0982 & \dashline & \textbf{-93.44\%**} & +39.55\%** & -0.35\%** & -17.44\%** & 2.4906 & \dashline & \textbf{-129.88\%**} & +82.08\%** & -0.52\%** & -29.79\%** \\
\hline
\midrule
\multicolumn{19}{c}{\textbf{Yelp}} \\
$LowPop$ & 0.0192 & \dashline & \textbf{+126.17\%**} & +12.65\% & +6.94\% & +6.94\% & 0.0040 & \dashline & \textbf{-63.60\%**} & -15.61\%** & -38.46\%** & -38.46\%** & 1.7702 & \dashline & \textbf{-101.15\%**} & -25.27\%** & -65.20\%** & -65.20\%** \\
$MedPop$ & 0.0304 & \dashline & -33.66\%** & \textbf{+2.22\%} & +1.15\% & -27.81\%** & 0.0079 & \dashline & \textbf{-75.92\%**} & -5.00\%** & -0.14\%** & -35.00\%** & 2.3983 & \dashline & \textbf{-107.50\%**} & -7.77\%** & -0.22\%** & -50.67\%** \\
$HighPop$ & \textbf{0.0650} & \dashline & -75.53\%** & -0.94\% & \textbf{+0.00\%} & -17.65\%** & 0.0093 & \dashline & \textbf{-74.34\%**} & +0.54\% & -0.01\% & -18.58\%** & 1.0542 & \dashline & \textbf{-145.19\%**} & +1.13\% & -0.04\% & -37.78\%** \\
$All$ & 0.0350 & \dashline & -31.70\%** & \textbf{+2.19\%} & +1.36\% & -20.24\%** & 0.0074 & \dashline & \textbf{-74.20\%**} & -4.74\%** & -4.21\%** & -31.24\%** & 2.0039 & \dashline & \textbf{-110.34\%**} & -9.93\%** & -11.68\%** & -51.88\%** \\
\hline

\end{tabular}
}
\end{table*} 

\paragraph{\textbf{Optimization.}} The optimization is guided by a trade-off parameter $\lambda \in [0, 1]$, controlling the balance between relevance $\text{Rel}(L_u)$ based on item scores from the recommendation model, and calibration measured by the Jensen-Shannon divergence $\Im(P, Q(L_u))$ between the user’s historical popularity distribution $P$ and the recommendation list's distribution $Q$. Higher divergence indicates a stronger mismatch between what the user prefers and what is recommended, resulting in a greater penalty. This is formally given by: 
\begin{equation} \label{eq:1}
L^*_u = \arg\max_{L_u, \, |L_u| = n} \left( (1 - \lambda) \cdot \text{Rel}(L_u) - \lambda \cdot \Im(P, Q(L_u)) \right)
\end{equation}
The final list $L^*_u$ is constructed iteratively via greedy optimization, where at each step, the item that maximizes the weighted trade-off between relevance and calibration is selected. 
To further personalize the debiasing procedure, we optimize $\lambda$ separately for each user group ($LowPop$, $MedPop$, and $HighPop$) via grid search, resulting in group-specific parameters that are inserted into Equation~\eqref{eq:1}. The optimal $\lambda$ values are found by:

%\begin{enumerate}
    %\item 
\vspace{1mm} \noindent \textbf{$CP_H$:} Maximizing the harmonic mean $H$ of accuracy (nDCG) and calibration (1 - $\Im$), following \citet{lesota_exploring_2022}: %, denoted as $CP_H$:
\begin{align}
    \lambda = \frac{\mathrm{nDCG}_g \cdot \left(1 - \Im_g(P, Q)\right)}{\mathrm{nDCG}_g + \left(1 - \Im_g(P, Q)\right)}
\end{align}
    
    %\item 
\vspace{1mm} \noindent \textbf{$CP_\Im$:} minimizing Jensen-Shannon divergence directly by setting $\lambda = 1$, prioritizing calibration over accuracy, denoted as $CP_\Im$.
%\end{enumerate}

\subsection{Training and Evaluation}
We generate baseline recommendations (BPR) using RecBole\footnote{\url{https://github.com/RUCAIBox/RecBole/tree/master}}~\citep{zhao_recbole_2021} and the context-aware recommendations (LORE and USG) using CAPRI\footnote{\url{https://github.com/CapriRecSys/CAPRI}}~\citep{tourani_capri_2024}. To foster reproducibility~\citep{semmelrock2025reproducibility}, our code and data samples are publicly available via GitHub\footnote{\url{https://github.com/andreafooo/POI_RS_PopBias_Mitigation}}. We train the models for 200 epochs, optimizing the learning rate, embedding size, and batch size on the validation sets. The top-150 recommendations per user are stored for $CP_H$ and $CP_\Im$ as a basis for re-ranking to evaluate the final top-10 recommendations. In this study, we evaluate accuracy via nDCG\citep{wang_theoretical_2013, jarvelin_cumulated_2002, jadon_comprehensive_2024}. 
The evolution of the recommendation popularity, and consequently, the level of popularity bias, is measured by average recommendation popularity (ARP)~\citep{mullner_impact_2024, abdollahpouri_user-centered_2021} and the popularity lift (PopLift)~\citep{mullner_impact_2024}. While ARP illustrates the popularity of the recommendations of each user, PopLift helps to reflect whether the popularity of the recommendations is higher, lower, or close to the popularity of items in the user profile for each user group.

\begin{table*}[t!]
\centering
\caption{RQ2. Evaluation results for LORE and USG algorithms combined with $CP_H$ and $CP_\Im$ methods. Results for LORE $Base$ and USG $Base$ are equivalent to LORE and USG from Table~\ref{tab:rq1} (best values are shown in bold); $\Delta$\% values per metric between $Base$ and $CP_H$/$CP_\Im$ in brackets, showing any significant t-test differences ($p < 0.05$ \textnormal{ as **}), Bonferroni-corrected for each metric.}
\label{tab:rq2}
\resizebox{\textwidth}{!}{%
\begin{tabular}{ll|llll|llll|llll}
\toprule
\multicolumn{1}{c}{\textbf{Model}} & \multicolumn{1}{c}{\textbf{Group}} & \multicolumn{4}{c}{\textbf{nDCG $\uparrow$ ($\Delta$\% nDCG)}} & \multicolumn{4}{c}{\textbf{ARP $\downarrow$ ($\Delta$\% ARP)}} & \multicolumn{4}{c}{\textbf{PopLift $\rightarrow 0$ ($\Delta$\% PopLift)}} \\
 &  & $Base$ & \dashline & $CP_H$ & $CP_\Im$ & $Base$ & \dashline & $CP_H$ & $CP_\Im$ & $Base$ & \dashline & $CP_H$ & $CP_\Im$\\
\hline
\midrule
\multicolumn{14}{c}{\textbf{Foursquare}} \\
LORE & $LowPop$ & 0.0172 & \dashline & \textbf{0.0255 (+48.51\%)} & \textbf{0.0255 (+48.51\%)} & \textbf{0.0069} & \dashline & 0.0119 (+72.26\%**) & 0.0119 (+72.26\%**) & -0.4715 & \dashline & \textbf{-0.1371 (+70.92\%**)} & \textbf{-0.1371 (+70.92\%**)} \\
LORE & $MedPop$ & 0.0122 & \dashline & \textbf{0.0212 (+74.07\%**)} & \textbf{0.0212 (+74.07\%**)} & \textbf{0.0064} & \dashline & 0.0132 (+107.75\%**) & 0.0132 (+107.75\%**) & -0.7920 & \dashline & \textbf{-0.5712 (+27.88\%**)} & \textbf{-0.5712 (+27.88\%**)} \\
LORE & $HighPop$ & 0.0056 & \dashline & \textbf{0.0200 (+258.16\%**)} & \textbf{0.0200 (+258.16\%**)} & \textbf{0.0062} & \dashline & 0.0149 (+139.93\%**) & 0.0149 (+139.93\%**) & -0.8740 & \dashline & \textbf{-0.6986 (+20.06\%**)} & \textbf{-0.6986 (+20.06\%**)} \\
LORE & $All$ & 0.0118 & \dashline & \textbf{0.0218 (+83.99\%**)} & \textbf{0.0218 (+83.99\%**)} & \textbf{0.0064} & \dashline & 0.0133 (+106.32\%**) & 0.0133 (+106.32\%**) & -0.7443 & \dashline & \textbf{-0.5099 (+31.50\%**)} & \textbf{-0.5099 (+31.50\%**)} \\
USG & $LowPop$ & 0.0224 & \dashline & \textbf{0.0228 (+1.72\%)} & 0.0222 (-1.03\%) & 0.1548 & \dashline & 0.1353 (-12.62\%**) & \textbf{0.1247 (-19.47\%**)} & 10.7837 & \dashline & 9.0601 (-15.98\%**) & \textbf{7.9484 (-26.29\%**)} \\
USG & $MedPop$ & \textbf{0.0956} & \dashline & 0.0937 (-2.03\%) & 0.0912 (-4.58\%) & 0.1340 & \dashline & 0.1287 (-3.90\%**) & \textbf{0.1189 (-11.26\%**)} & 3.4161 & \dashline & 3.2405 (-5.14\%) & \textbf{2.9022 (-15.04\%**)} \\
USG & $HighPop$ & \textbf{0.1620} & \dashline & 0.1602 (-1.12\%) & 0.1552 (-4.15\%) & 0.1286 & \dashline & 0.1253 (-2.56\%) & \textbf{0.1168 (-9.19\%**)} & 1.6424 & \dashline & 1.5728 (-4.24\%) & \textbf{1.3966 (-14.97\%**)} \\
USG & $All$ & \textbf{0.0942} & \dashline & 0.0928 (-1.54\%) & 0.0902 (-4.27\%) & 0.1371 & \dashline & 0.1294 (-5.62\%**) & \textbf{0.1196 (-12.72\%**)} & 4.5349 & \dashline & 4.0709 (-10.23\%**) & \textbf{3.6103 (-20.39\%**)} \\
\hline
\midrule
\multicolumn{14}{c}{\textbf{Yelp}} \\
LORE & $LowPop$ & \textbf{0.0434} & \dashline & 0.0426 (-1.69\%) & 0.0390 (-10.03\%) & \textbf{0.0014} & \dashline & \textbf{0.0014 (+0.00\%)} & 0.0015 (+5.37\%) & -0.0203 & \dashline & -0.0210 (-3.66\%) & \textbf{0.0052 (+125.79\%)} \\
LORE & $MedPop$ & 0.0201 & \dashline & \textbf{0.0254 (+26.22\%)} & \textbf{0.0254 (+26.22\%)} & \textbf{0.0019} & \dashline & 0.0022 (+16.65\%**) & 0.0022 (+16.65\%**) & -0.1798 & \dashline & \textbf{-0.0473 (+73.71\%**)} & \textbf{-0.0473 (+73.71\%**)}\\
LORE & $HighPop$ & 0.0159 & \dashline & \textbf{0.0269 (+68.99\%)} & \textbf{0.0269 (+68.99\%)} & \textbf{0.0024} & \dashline & 0.0032 (+34.00\%**) & 0.0032 (+34.00\%**) & -0.4764 & \dashline & -0.3043 (-36.12\%**) & -0.3043 (-36.12\%**) \\
LORE & $All$ & 0.0239 & \dashline & \textbf{0.0292 (+21.79\%)} & 0.0284 (+18.77\%) & \textbf{0.0019} & \dashline & 0.0023 (+18.46\%**) & 0.0023 (+19.29\%**) & -0.2072 & \dashline & -0.0934 (+54.91\%**) & \textbf{-0.0882 (+57.44\%**)} \\
USG & $LowPop$ & 0.0216 & \dashline & \textbf{0.0246 (+14.09\%)} & \textbf{0.0246 (+14.09\%)} & 0.0033 & \dashline & \textbf{0.0022 (-35.61\%**)} & \textbf{0.0022 (-35.61\%**)} & 1.3230 & \dashline & \textbf{0.4271 (-67.72\%**)} & \textbf{0.4271 (-67.72\%**)} \\
USG & $MedPop$ & 0.0310 & \dashline & \textbf{0.0315 (+1.58\%)} & 0.0274 (-11.89\%) & 0.0075 & \dashline & 0.0075 (+0.00\%) & \textbf{0.0048 (-35.75\%**)} & 2.2118 & \dashline & 2.2066 (-0.23\%) & \textbf{1.0415 (-52.91\%**)} \\
USG & $HighPop$ & \textbf{0.0644} & \dashline & \textbf{0.0644 (+0.00\%)} & 0.0496 (-22.87\%) & 0.0094 & \dashline & 0.0093 (-0.26\%) & \textbf{0.0075 (-19.45\%**)} & 1.0661 & \dashline & 1.0605 (-0.53\%) & \textbf{0.6456 (-39.44\%**)} \\
USG & $All$ & 0.0358 & \dashline & \textbf{0.0367 (+2.52\%)} & 0.0313 (-12.70\%) & 0.0070 & \dashline & 0.0068 (-3.55\%) & \textbf{0.0048 (-31.41\%**)} & 1.8049 & \dashline & 1.6215 (-10.16\%**) & \textbf{0.8395 (-53.49\%**)} \\
\hline

\end{tabular}
}
\end{table*}

%%%%%%%%%%%%%%%%%%%%
\section{Results}
In this section, we present our results according to our two research questions. As mentioned beforehand, for the sake of space, we only discuss our findings for the Foursquare and Yelp datasets. 

\paragraph{\textbf{RQ1. Independent Impact of Context-Awareness and Popularity Calibration.}} 
We depict our results for RQ1 in Table~\ref{tab:rq1} and evaluate how the context-aware recommendation models LORE and USG, as well as popularity calibration, (1) optimized for the weighted mean between nDCG and $\Im$ ($CP_H$), and (2) optimized to minimize $\Im$ ($CP_\Im$), perform independently compared to the non-contextualized BPR baseline. 

\begin{figure*}[t]
  \centering
  \begin{subfigure}[b]{0.80\textwidth}
    \centering
    \includegraphics[width=\textwidth]{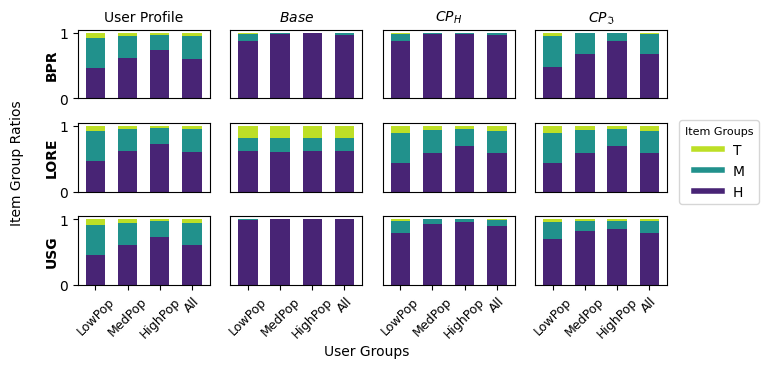}
    \caption{Foursquare}
    \Description{Stacked bar charts showing the distribution of items (T, M, H) for the user groups (LowPop, MedPop, HighPop) in their user profile vs. BPR, LORE and USG $Base$, and the respective $CP_H$, and $CP_\Im$ in Foursquare, showing a strong bias for H-items in BPR $Base$ and a broader mix of items yet a lack of distributional approximation in LORE $Base$, approximation for LowPop in BPR $CP_H$, approximation for all groups in BPR $CP_\Im$ except for T-items, and close approximation in LORE $CP_H$ and $CP_\Im$ including T-items.}
  \end{subfigure}

  \vspace{2mm}

  \begin{subfigure}[b]{0.80\textwidth}
    \centering
    \includegraphics[width=\textwidth]{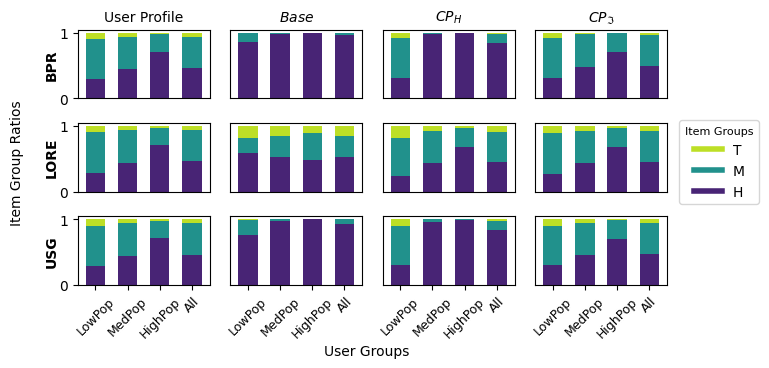}
    \caption{Yelp}
    \Description{Stacked bar charts showing the distribution of items (T, M, H) for the user groups (LowPop, MedPop, HighPop) in their user profile vs. BPR, LORE and USG $Base$, and the respective $CP_H$, and $CP_\Im$ in Yelp, showing a strong bias for H-items in BPR $Base$ and a broader mix of items yet a lack of distributional approximation in LORE $Base$, approximation for LowPop in BPR $CP_H$, approximation for all groups in BPR $CP_\Im$ except for T-items, and close approximation in LORE $CP_H$ and $CP_\Im$ including T-items.}
  \end{subfigure}

  \caption{RQ1 \& RQ2. Comparing the item group ratios (T, M, H) in the groups' user profiles to BPR, LORE, and USG $Base$, $CP_H$, and $CP_\Im$ in (a) Foursquare and (b) Yelp.}
  \label{fig:popularity_distribution}
  \vspace{-3mm}
\end{figure*}

The impact of context-awareness depends on the chosen model and dataset\footnote{For full results on all datasets, please see: \url{https://github.com/andreafooo/POI_RS_PopBias_Mitigation}}. LORE significantly lowers ARP and PopLift for all user groups in both datasets, effectively diminishing popularity bias and delivering niche recommendations, but often at the cost of lower accuracy, except for niche user groups like $LowPop$, where accuracy is significantly improved in the Yelp dataset. Figure~\ref{fig:popularity_distribution} shows that T-items increase in all groups in LORE $Base$ compared to the user profile. USG has a contrary effect in Foursquare, producing recommendations that almost exclusively contain H-items, leading to a significant increase in ARP and PopLift for all user groups, thus increasing the level of popularity bias, compared to the general baseline.
Simultaneously, accuracy decreases significantly for all user groups, except $HighPop$, where the decrease is non-significant. The context-aware algorithm leads to better results on Yelp with a significant decline in ARP and PopLift and a non-significant increase in nDCG for all user groups except $HighPop$.

$CP_H$ leads to a significant decrease in ARP and PopLift for all user groups except $LowPop$ in Foursquare, whilst neither significantly impacting accuracy, nor approximating the user profile distribution as illustrated in Figure~\ref{fig:popularity_distribution}. In the Yelp dataset, no significant changes are observed using the accuracy-oriented calibration technique, except for $HighPop$. This can also be seen in Figure~\ref{fig:popularity_distribution}, where BPR $Base$ is heavily biased towards the most popular H-items, while $CP_H$ approximates the distribution for $LowPop$ users and increases accuracy, yet fails to do so for the other groups. $CP_\Im$, optimized to align the popularity of the recommendations with the user profile, has more profound effects on both accuracy and popularity bias, leading to significant decreases in accuracy and popularity bias for all users in the Foursquare dataset, except for the $LowPop$ users' accuracy. In Yelp, a significant decline in ARP and PopLift, hence a reduced level of popularity bias, is evident for all user groups. However, this also impacts accuracy for all groups, except $LowPop$, where a non-significant increase in nDCG is achieved after calibration. $CP_\Im$ approximates the distribution of H- and M-items to the user profiles, yet T-items remain underrepresented (see Figure~\ref{fig:popularity_distribution}). 

In the PopLift metric, the baseline and calibrated BPR recommendations and USG produce recommendations that exceed the user profile's popularity level. In contrast, LORE delivers recommendations below the user profile's popularity level, yielding the highest accuracy for $LowPop$ users, also suggesting potential for better personalization among underrepresented users. 

%%%%%%%%%%%%%%%%%%%%%%%%%%%%%%%%%%%%%%%%%%%
\paragraph{\textbf{RQ2. Combined Impact of Context-Awareness and Popularity Calibration.}}  
To address RQ2, we evaluate the effects of combining context-aware models USG and LORE with $CP_H$ and $CP_\Im$ against their respective $Base$ (top-10 recommendations without calibration) results, and depict them in Table \ref{tab:rq2}. 

LORE ($Base$) recommends less popular items than the user profile, as highlighted by the negative PopLift values in Foursquare and Yelp, or the high proportion of T-items in Figure~\ref{fig:popularity_distribution}, yet lacks distributional similarity to the user profiles. Applying CP-based re-ranking to LORE improves accuracy in both datasets for all groups except the $LowPop$ user group in Yelp, but also increases the ARP of the recommended items, thereby reducing diversity. Nonetheless, our analysis of the PopLift metric suggests that, despite this increase in popularity, the combination of LORE with CP aligns the popularity distribution of the recommendations more closely with the user profile than other methods and combinations, also managing to include T-items in the recommendation lists of all user groups, as opposed to applying CP to BPR.

USG ($Base$), in combination with the accuracy-oriented $CP_H$, shows no significant impact on nDCG in either dataset. We generally observe a positive, yet non-significant, effect on the $LowPop$ accuracy in both datasets. Concerning popularity bias, in Foursquare, the combination significantly reduces ARP for all groups except $HighPop$, and improves PopLift significantly for $LowPop$ and All. However, as illustrated in Table~\ref{tab:rq1}, USG $Base$ exhibits a particularly high level of popularity bias in Foursquare, creating recommendations more than 10 times as high as the user profile for $LowPop$ users, vs. more than 4 times as high for BPR. In Yelp, where BPR and USG yield similar results, combining USG and $CP_H$ significantly reduces both popularity bias metrics for $LowPop$, while leading to equal levels or non-significant increases in accuracy for the user groups. The combination of USG with $CP_\Im$ consistently reduces ARP and PopLift across all user groups in both datasets. These reductions are statistically significant, indicating popularity bias mitigation while maintaining accuracy and reducing bias.

%%%%%%%%%%%%%%%%%%%%
\section{Discussion and Conclusion}
In this paper, we evaluated the effectiveness of context-awareness and popularity calibration to mitigate popularity bias in POI recommendations. In general, we find that the effectiveness of context-awareness varies substantially between models, while the effectiveness of CP can vary for different user groups and is dictated by the chosen $\lambda$ that balances accuracy and calibration. The combination of context-awareness and CP helps counteract the shortcomings of both methods when applied individually. 
%, resulting in recommendations that better reflect the user's user profile and include niche items for all user groups, where applying CP to a general baseline failed to do so.

The two context-aware models yield contrasting results, with LORE producing the most niche recommendations and USG producing the most biased recommendations in all datasets except Yelp. In all four datasets, the combination of BPR and $CP_H$ hardly improves the biased item distribution and popularity bias, especially for $MedPop$ and $HighPop$ users, whereas $CP_\Im$ approximates the distribution in the user profile, except for T-items that remain underrepresented, suggesting a lack of T-items in the top-150 recommendations produced by BPR that form the basis for the re-ranking (RQ1). 
Combining context-aware models and CP, especially LORE, increases the accuracy and popularity of recommendations, while recommending slightly more T-items than the user profiles. This can closely align recommendations with the user profiles' item distributions and PopLift values, potentially counteracting popularity bias over time. In our study, the combination of LORE and CP is the only method that accurately reflects the distribution of T-items, suggesting that the combination can counteract the algorithms' low user fairness that was discussed in previous research \citep{rahmani_unfairness_2022} (RQ2). Therefore, if mitigating popularity bias is the primary objective, our findings suggest combining LORE with CP. Conversely, if preserving accuracy while slightly reducing bias is preferred, general models such as BPR or context-aware models like USG combined with CP and tuned towards accuracy (e.g., $CP_H$) are more suitable.  

%\paragraph{\textbf{Future Work.}} 
To validate these findings and assess their practical relevance, in our future work, we plan to investigate the effects of context-aware and calibration-based combinations in user studies, particularly focusing on their impact on user satisfaction and perceived recommendation quality across different user groups. In addition, we plan to study how these methods impact other RS stakeholders, in particular POI providers, as well, by following principles of multistakeholder RS evaluation~\citep{dagstuhl_multi_2024,burke2025centering}. 

\begin{acks} 
This research was supported by the Austrian FFG COMET program and the FFG Femtech project Radreisen4All. 
\end{acks}

% Future work should focus on online studies within active systems to evaluate user preferences, particularly for niche models like LORE, potentially in conjunction with CP, and assess if its accuracy remains low. Additionally, adopting user-group-specific model choices could help improve fairness and effectiveness in recommendations.

% \begin{acks}

% \end{acks}

\balance
\bibliographystyle{ACM-Reference-Format}
\bibliography{main.bib}

\end{document}